\documentclass[preprint,10pt,a4paper]{aastex}
\usepackage{amsmath}
\usepackage{graphicx}
\usepackage{multicol}
\newcommand{\be}{\begin{equation}} 
\newcommand{\ee}{\end{equation}}
\newcommand{\nn}{\mbox{} \nonumber \\ \mbox{} }
\newcommand{\ba}{\begin{eqnarray}}
\newcommand{\ea}{\end{eqnarray}}

\newcommand{\Alfven}{ Alfv\'{e}n }

\newcommand\etal{\textit{et al.\ }}
\newcommand\eg{\textit{e.g.,}}

\newcommand{\Bf}{{magnetic field}}
\newcommand{\Bfs}{{magnetic fields}}

\newcommand{\NS}{neutron star}
\newcommand{\NSs}{{neutron stars}}

\newcommand{\ms}{magnetosphere}
\newcommand{\mss}{magnetospheres}

\begin{document}

\title{Magnetospheric ``anti-glitches'' in magnetars}
\author{Maxim Lyutikov\\
Department of Physics, Purdue University, 
 525 Northwestern Avenue,
West Lafayette, IN \\
The Canadian Institute for Theoretical Astrophysics,
University of Toronto, 
60 St. George Street 
Toronto, Ontario, M5S 3H8
Canada
}

\begin{abstract} 
We attribute the rapid spindown of  magnetar  1E 2259+586 observed by Archibald \etal\ (2013), termed the ``anti-glitch'',  to partial opening of the \ms\ during the $X$-ray burst,  followed by 
 changes of the structure of the  closed field line region.  To account  for the observed spin decrease during the $X$-ray flare all that is needed is the transient  opening, for just  one period, of a  relatively  small fraction of the \ms, of the order of only few percent.  More generally, we argue that in magnetars all timing irregularities have magnetospheric origin and are induced either by (i) the fluctuations in the current structure of the \ms\ (similar to the  long term torque variations in the rotationally powered pulsars); or, specifically to magnetars,  by (ii) opening of a  fraction of the \ms\ during bursts and flares - the latter events are always  accompanied by rapid spindown, an ``anti-glitch''.
 Slow rotational  motion of the \NS\ crust, driven by crustal \Bfs,   leads  beyond some twist  limit  to explosive instability   of the  external  \Bfs\ and  transient  opening of  a large magnetic flux in  a CME-type event,  the  post-flares increase of  magnetospheric currents accompanied by enhanced $X$-luminosity and  spindown rate, changing profiles,   as well as  spectral hardening, all in agreement with the magnetospheric model of torque fluctuations.

\end{abstract}

\section{Magnetars' bursts and flares}

Magnetar emission \citep{2006csxs.book..547W} is powered by dissipation of a non-potential (current-carrying) magnetic field \citep{tlk}. The magnetic  field exerts Lorentz force on the crust, which is balanced by induced elastic stress.  For strong enough magnetic fields, Lorentz force may induce a stress that exceeds the critical stress of the lattice, leading to crustal failure. 
This leads to lattice failure of the crust \citep{2011arXiv1109.5095H}, which can be either in a form of fast propagating crack \citep{TD95} (and   sudden 
untwisting of the internal magnetic
field) or a slowly (plastically)  propagating fault  \citep{2003ApJ...595..342J} that leads to slow evolution of the external  \Bfs\  \citep{lyut06}. In the latter case  twisting of the external \Bfs\   leads to   a 
 sudden relaxation of the  twist outside  of the
star   in  analogy with solar flares and Coronal Mass Ejections (CMEs)  \citep{ForbesCME,lyut06,2012ApJ...754L..12P}. Perhaps the best argument in favor of magnetic storage of energy  outside of the star is a very short rise time of the 2004 giant flares,  $\sim 0.25 $ msec
\citep{palmer}. This  points to the
magnetospheric origin of giant flares \citep{lyut06}.

\cite{tlk}  developed a model of the emission and spin-down behavior of the
Anomalous X-ray Pulsars (AXPs) and Soft Gamma-ray Repeaters (SGRs)
in which the
 dissipation of magnetic fields 
powers its unusually bright X-ray luminosity.
In this model the magnetic fields
anchored in a well conducting   stellar   interior are dissipated in  the  magnetosphere. \cite{1995ApJ...443..810W,tlk,2009MNRAS.395..753P}    constructed  a self-similar
force-free current-carrying magnetosphere in which the 
 current twist the field lines and causes them to expand with
 respect to a potential dipole field. Expansion of field lines results in 
a changing torque  on the neutron star which is observed as a variable
spin-down rate. Since the  X-ray luminosity is proportional to the 
strength of the currents flowing in the magnetosphere,  one 
expects that spin-down rate is positively correlated with luminosity.

\cite{lyut06} (see also \cite{2012ApJ...754L..12P}) proposed that magnetar flares are magnetospheric  (and {\it not} crustal) events similar to the Solar flares and CMEs. According to \cite{lyut06}
magnetars bursts and flares are 
 driven by  unwinding  of the internal non-potential magnetic field
 which leads to 
 a slow build-up of magnetic energy  outside of the neutron star. For large magnetospheric currents,
 corresponding to 
 a large twist of the  external  magnetic field,
 magnetosphere  becomes dynamically unstable on the \Alfven crossing times scale
 of the inner magnetosphere. The reconnection processes, \eg\ the tearing mode,  in the strongly magnetized  plasma of magnetar \mss\  develops similarly to the non- relativistic plasmas \citep{LyutikovTear,2007MNRAS.374..415K}.

 In addition, current carrying charges
 resonantly scatter thermal photons from the surface producing 
through multiple  resonant scattering (resonant Comptonization)
a non-thermal  power-law  X-ray tail \citep{2006MNRAS.368..690L,2007ApJ...660..615F,2008ApJ...686.1245R,2013ApJ...762...13B}. The hardness of the power-law 
depends on the typical velocity of the current carrying particles
and their density.  For larger densities (and larger currents)
a given photon has a larger probability to be scattered at the cyclotron 
resonance and thus will experience greater amount of scattering, gaining 
energy stochastically  in each scattering 
through Doppler effect.
Thus, there is a natural prediction of the model by  \cite{2006MNRAS.368..690L}, that the
harness of the  power-law  X-ray tail should correlate 
positively with the spin down rate and the $X$-ray luminosity.

As we discuss below, the observations of the rapid spindown of the magnetar  1E 2259+586 \citep{2013Natur.497..591A} are in general agreement with the  magnetospheric model.  Thus, they   are qualitatively different form  glitches in the rotationally powered pulsar, where they are driven by the presence of a faster component inside the \NS.

\section{Observations of the  ``anti-glitch'' in magnetar  1E 2259+586}

 \cite{2013Natur.497..591A} reported a fast change of the rotation frequency of the magnetar  1E 2259+586, which occurred within  two weeks of a bright $X$-ray flare. The accumulated change in frequency was $ \Delta \nu \sim10^{-7} $ Hz over two weeks. For approximately a few months after the flare both  the flux and the spindown rate  were about two times larger than normal. In addition, the flux   increase was also accompanied by hardening of the spectrum and a change in the pulse profile. 
 Importantly, there was a GBM burst, presumably associated with the glitch, with peak luminosity of the order $10^{38}$ erg s$^{-1}$ and duration of $\sim 100$ milliseconds.
 
By analogy with glitches in radio pulsar, except, importantly, for the different sing of the period change, the preferred interpretation of  \cite{2013Natur.497..591A} for the rapid spindown is  a  glitch internal to the \NS. This requires a \NS\ component that rotates slower than the crust. The origin of such a component is not obvious \citep[see, though, ][]{2000ApJ...543..340T}.

In radio pulsars the notion of a glitch is usually referred to  a sudden change of the spin frequency, on time scale comparable to the pulsar period. In magnetars,   the timing measurements are usually separated by about a week. In case of   1E 2259+586, the anti-glitch could be a sudden event, on the time scale of rotation, as well due to enhanced spindown rate by an order of magnitude over approximately a week. 
 
Below we discuss  an interpretation alternative  to that of  \cite{2013Natur.497..591A}. We argue that all the  spin-down activity, as well as spectrum hardening, flux increase  and profile changes during the activity period,  can be explained via purely magnetospheric effects within the twisted \ms\ model of \cite{tlk} described above, plus an additional effect of opening the \ms\ during the prompt burst \citep{lyut06}. Thus, we envision that  a large spindown occurred during an  opening of a modest fraction of the \ms,  on a time scale as short as  one rotation,  followed by  an extended period of high spin-down rate, driven by the magnetospheric currents.

\section{Twisted Force-Free Equilibria}
\label{force-free}

In this Section we review the solution of \cite{1995ApJ...443..810W,tlk} (see also \cite{2008MNRAS.385..875G,2009MNRAS.395..753P}), giving a few simple approximations that are later used for numerical estimates.
In a static force-free approximation the electric current flows along \Bf,
\be
{\bf\nabla}\times{\bf B} = \alpha({\cal P}) {\bf B}.
\label{ff}
\ee
 The non-rotating magnetic equilibria are then described by the second-order elliptic differential 
equation, the Grad-Shafranov equation  \citep{1966RvPP....2..103S}:
\be
{\bf \nabla} \left( {1\over r^2 \sin ^2 \theta} {\bf \nabla} {\cal P} \right) +
{1 \over  r^2 \sin ^2 \theta} I {  d I \over d {\cal P}} =0 
\label{GS}
\ee
 where poloidal and toroidal fields are 
 \begin{equation}\label{bpval}
{\bf B}_P = {{\bf\nabla}{\cal P} \times {\hat\phi}\over  r\sin\theta},
 \,
{\bf B}_T = - { 2 I \over r\sin\theta} {\hat\phi}
\ee
$I=I({\cal P})$  is the  total current enclosed by the magnetic flux
surfaces ${\cal P}$.
Choosing separable solutions, 
and assuming that the  current function is 
a power-law, $I \sim  {\cal P}^{(1+p)/p} $, Eq. (\ref{GS})  becomes
\be
r^2 { {\cal R}^{\prime \prime}_{rr} \over {\cal R}}+
(1-\mu^2) { F ^{\prime \prime}_{ \mu \mu}  \over F} 
+ C {r^2 \over r_\ast^2} \left( {\cal R} F \right)^{2/p}=0
\label{ss}
\ee
where $C$ is some dimensionless 
constant related to the strength of the currents
($I\propto \sqrt{ C p /(p+1)} {\cal P}^{(1+p)/p}$) and $r_\ast$ is a radius of a
neutron star.
Separable solutions of 
Eq.  (\ref{ss}) satisfy
\ba &&
 {\cal R} \sim \tilde{r}^{-p}
\nn &&
p(p+1)F + (1-\mu^2)
{\partial^2 F\over\partial\mu^2} = -CF^{(p+2)/p}
\label{gseq}
\ea
where $ \tilde{r}= r/r_\ast$ has been introduced.

The current-free case  comprises  the monopole solution $p=0$ (with an equatorial
 current sheet),
and vacuum  multipolar solutions  $p=1,2,....$. 
For non-integer $p$ system (\ref{gseq}) gives  current-carrying solutions
of type considered by \cite{1994MNRAS.267..146L}.
Magnetic fields and current densities are then given by
\ba &&
{B_r \over B_\ast} = - {F'(\mu) \over 2 \tilde{r}^{p+2}}
\nn &&
{B_\theta \over B_\ast} = { p F \over  2 \sqrt{1-\mu^2}  \tilde{r}^{p+2}}
\nn &&
{B_\phi\over B_\theta} = -
\left[{C\over p(1+p)}\right]^{1/2}\,F^{1/p}
\nn &&
j_\phi  { c r_\ast \over 4 \pi  B_\ast} =
 - { C F^{(p+2)/p} \over \sqrt{1-\mu^2} r^{(3+p)}}
\nn &&
j_\theta  { c r_\ast \over 4 \pi  B_\ast}=
- \sqrt{ { p C \over p+1}}  { p F^{(p+1)/p} \over r^{(p+2)}  \sqrt{1-\mu^2} }
\nn &&
{ j_r } { c r_\ast \over 4 \pi  B_\ast} =  -
 \sqrt{ {  C (1+p) \over p}} { 1 \over \tilde{r}^{(p+2)} }
F ^{1/p} F'
\label{sol}
\ea
Comparing with equation (\ref{ff}) shows that $\alpha({\cal P})$
is proportional to ${\cal P}^{1/p}$, and on dimensional grounds
one can write
\be
\alpha({\cal P}) = {C^{1/2}\over r_{\rm \ast}}\left({p+1\over p}\right)^{1/2}\,
\left({{\cal P}\over{\cal P}_0}\right)^{1/p}
\ee

  The poloidal components of
Eq. (\ref{ff}) can be integrated to give
\begin{equation}\label{bphival}
B_\phi = {\int \alpha({\cal P})\,d{\cal P}\over r \sin\theta}
= {p\over p+1}\,{{\cal P}\alpha({\cal P})\over  r \sin\theta}.
\end{equation}
Substituting Eqs. (\ref{bpval})-(\ref{bphival}) into the $\phi$-component
of Eq. (\ref{ff}) then gives the non-linear equation (\ref{gseq})
for the angular function $F = F(\mu)$.

The solution of Eq. (\ref{gseq}), including the dependence $p(C)$,
is uniquely defined by the parameter $C$
 and by the {\it three} boundary
conditions.  The first
 is the requirement of  zero  $B_{\phi}$ at the polar axis
(absence of line current):
$ B_{\phi}(\mu=1) \propto F(1)=0$ (nonzero  $B_{\phi}$ at the polar axis 
implies a line current), another is overall normalization
of the field which we take to be  $F' = {\rm const}
= -2$ at $\mu = 1$ (corresponding to a fixed flux density $B_{\rm \ast}$
at the magnetic pole). The third boundary condition
specifies  the multipole structure of the magnetic field at small
currents.
For dipole field   $B_R \propto F' = 0$ at
$\mu  = 0$, for quadrupole field 
  $B_{\theta}  \propto F = 0$ at
$\mu  = 0$.

In the absence of current, $C=0$, Eq. (\ref{gseq}) 
with a boundary condition $F(1)=0$ 
has  solutions of a 
 vacuum multipole field,  
e.g. dipole  field for $p=1$ and  $F \propto 1-\mu^2 = \sin^2 \theta$, 
or quadrupole for  $p=2$ and $ F \propto \mu (1-\mu^2)$. 
Interestingly, there is also a
  monopole solution, $p=0$, $F \propto (1-\mu)$.
For nonzero current  Eq. (\ref{gseq}) has to be solved numerically.

Presence of a current leads to  twisting of  magnetic field lines.
A magnetic field line anchored at polar angle $\theta$ will twist 
through a net angle
\begin{equation}\label{deltaphi}
\Delta\phi(\theta) = \int 
{B_\phi\over B_\theta} {d\theta\over\sin\theta} =
 -
\left[{C\over p(1+p)}\right]^{1/2}\, \int F^{1/p} { d \mu \over 1- \mu^2}
\end{equation}
(integration is between the anchor points of a given field line)
before returning to the stellar surface.
Twisting of fields imply a net helicity 
\begin{equation}\label{helicity}
{\cal H}_B = \int {\bf A}\cdot{\bf B}\,dV =
{3\pi\over 2}\,(B_{\rm \ast}R_{\rm \ast}^2)^2\,\sqrt{C\over p(p+1)}\,
\int d\mu {[F(\mu)]^{2+1/p}\over 1-\mu^2}.
\end{equation}
where ${\bf A}$ is vector potential.

Equation for flux surfaces can be derived from the
equation $  d \ln \tilde{r}/d \theta = B_r/B_\theta$ which gives
\be
 \tilde{r}=\left( {F \over F_\ast}\right)^{1/p}
\ee

For nonzero current Eq. (\ref{gseq}) has to be solved numerically
for functions $F(\mu)$ and $p(C)$,
Figs. \ref{Fofmu}, \ref{PofC}, \ref{phiSN}.
For each value of $C\leq 0.873$ there
are two solutions for $p$.  The upper branch connects continuously to the 
vacuum dipole $C = 0$, $p = 1$ ($F=1-\mu^2$, 
$B_r, B_\theta \propto /\tilde{r}^{-3}$), Eq. (\ref{small}); and 
the lower branch connects to  a   
split monopole $C = 0$, $p = 0$ configuration
($F=2(1-\mu)$, $B_r = B_{\ast} /\tilde{r}^2 $,  Appendix  \ref{p=0}).

\begin{figure}
\includegraphics[width=.99\linewidth]{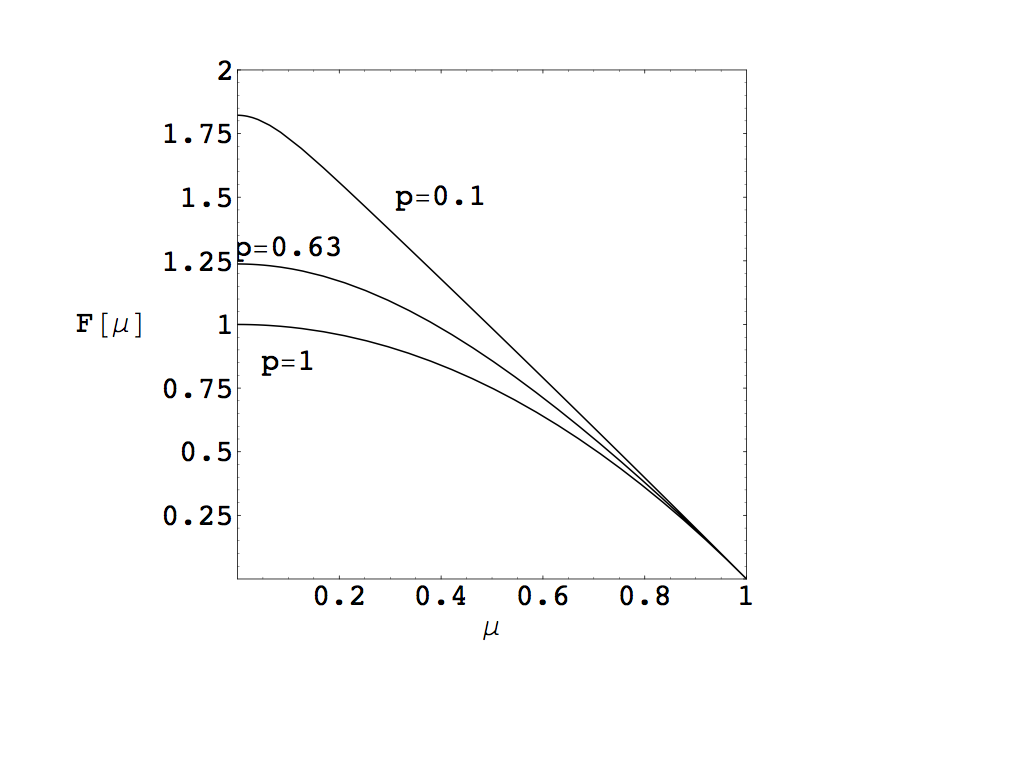}
\caption{
Function $F(\mu)$ for different values of $p$. For $p=0$, $F=2(1-\mu)$ almost everywhere except close to $\mu =0$ point, see Appendix \ref{p=0}}
\label{Fofmu}
\end{figure}

\begin{figure}
\includegraphics[width=.99\linewidth]{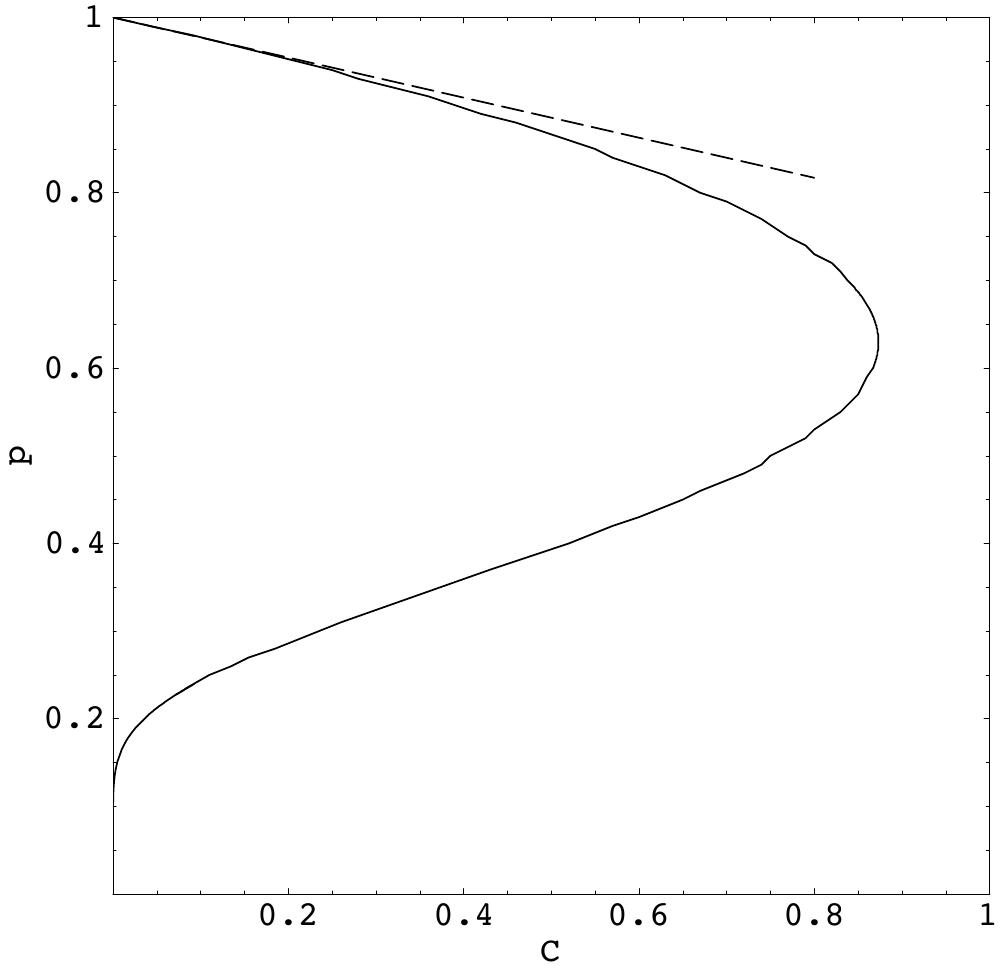}
\caption{
Function $p(C)$. The extremum is at $C=0.87$, $p=0.63$. 
Dashed line is a small twist approximation (Eq. (\ref{small}))}
\label{PofC}
\end{figure}

\begin{figure}
\includegraphics[width=.99\linewidth]{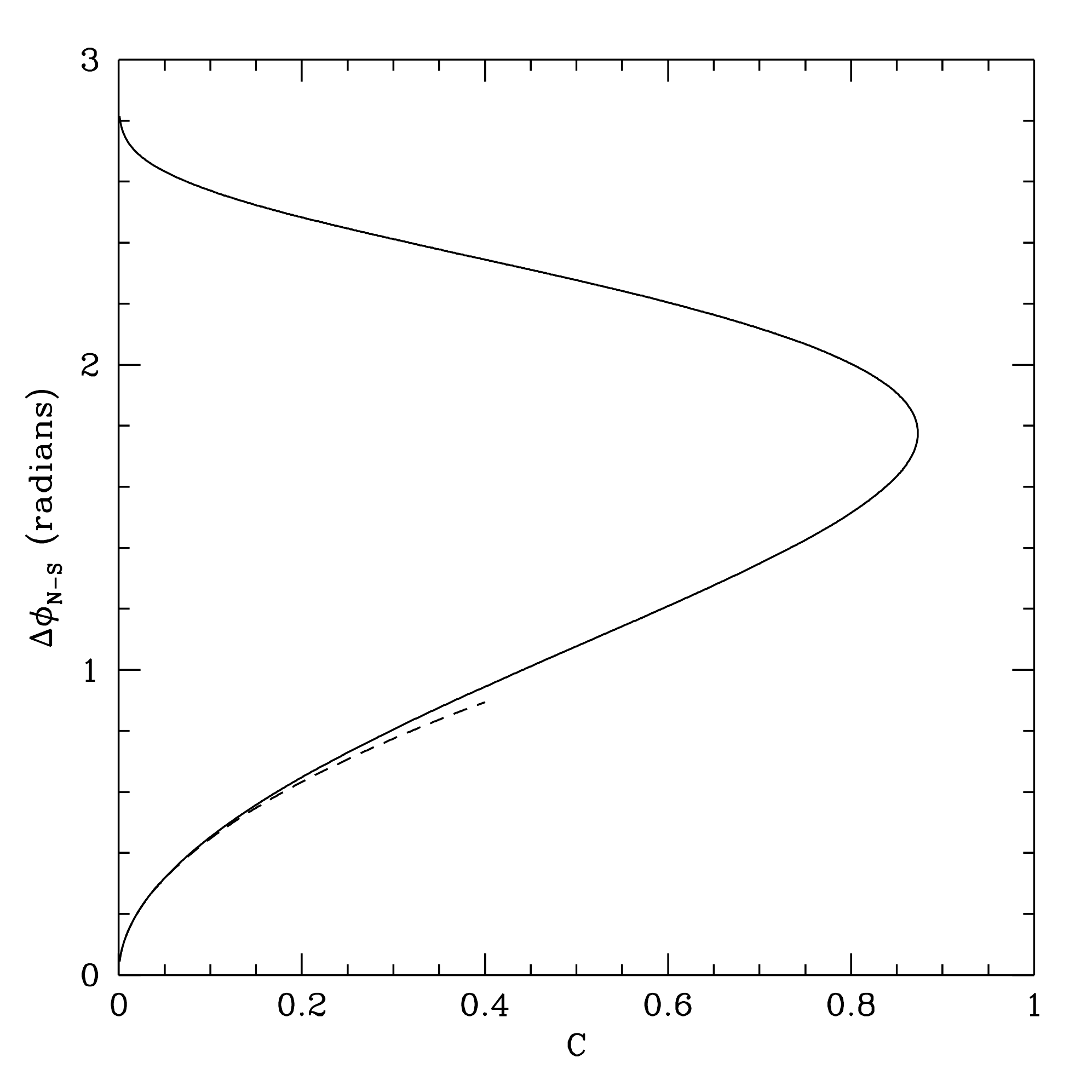}
\caption{ The  net twist angle between two poles
 versus current parameter $C$. 
Dotted line is the small current limit (Eq. (\ref{small}))}
\label{phiSN}
\end{figure}

As the current increases the distribution of the poloidal magnetic flux
evolves towards more spherical configuration, 
becoming  isotropic  in the limit $p=0$ (Fig. \ref{surf}). 
At the same time the distribution of a radial  current  at the surface
evolves from an approximately isotropic to strongly  concentrated
near equator for small $p$.
The field lines become more twisted, but for small  $p \rightarrow 0$ the twist
is confined to narrow layers near the equator, while the rest of the field
lines have little twist and also are straitened out 
(as compared with dipole) in the radial direction.
A given field line has  largest twist where it is furthest from the
star - at the equator. 

The radial dependence of the magnetic field softens to
${\bf B} \propto r^{-2.87}$ when $\Delta\phi_{\rm N-S} = 1$ radian.  The net
twist approaches $\Delta\phi_{\rm N-S} = \pi$ (one-half turn) in 
the split monopole limit ($p = 0$).  For comparison, a twisted 
quadrupole
 magnetic field  expands to infinity
after $1/\sqrt{3}$ turns \citep{1994MNRAS.267..146L}.
In the limit $p=0$ the field structure approaches split monopole:
$ F= 2(1-\mu)$, with field line  straightened out in the 
radial direction,
$B_r/B_\ast = {\rm sign[\mu]}/\tilde{r}^2$, $B_\theta =0$ plus 
 a current sheet 
in the equatorial plane which curries a surface current
$g_\phi= 2 {B_\ast} c /( 2  \pi \tilde{r}^2) $.
 The magnetic flux leaves the northern hemisphere
and after reaching infinity returns to the southern hemisphere.

\begin{figure}
\includegraphics[width=.99\linewidth]{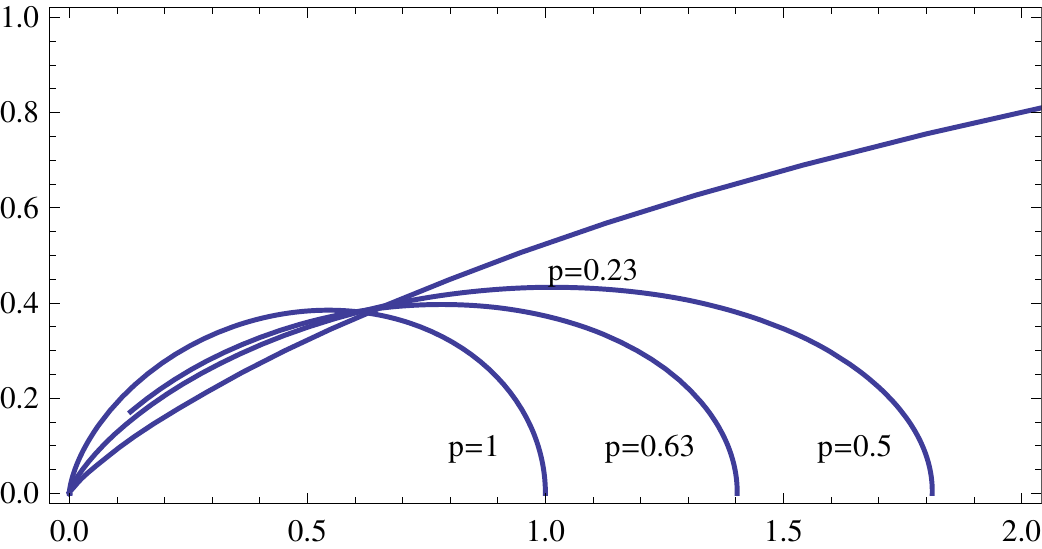}
\caption{
Form of the magnetic flux surfaces for different $p$; $p=1$ case corresponds to a dipole, $p=0$ is a monopole with radial poloidal fields.}
\label{surf}
\end{figure}

The total current flowing through one hemisphere at radius $\tilde{r}$
is given
by integral of $j_r$,
\be 
I = { 8 \pi^2  e B_\ast r_\ast  \over c } \, \sqrt{ C p \over p+1} \, 
{1 \over  \tilde{r}^p} \,  F(0)^{(p+1)/p)}.
\ee
Thus, a current that reaches to radius $ \tilde{r}$ for a given twist
decreases as $1/ \tilde{r}^p$. 
On the other hand, the current flowing through the star (at $\tilde{r}=1$)
increases continuously with decreasing $p$ and reaches
$I_{ \rm max} = 8 pi^2 B_\ast r_\ast /c $ in the limit
$p \rightarrow 0$. In fact, in this limit both
northern (outgoing) and southern (ingoing) poloidal 
currents are confined to the
equatorial plane and eliminate each other, leaving only toroidal current.

For small twists we can solve Eq. (\ref{gseq}) 
analytically by expanding near $p = 1,\, C  = 0 $ and $F= 1 - \mu^2$.
We find then $ p = 1 - 8  C/35$ and
\be
F =
(1- \mu^2) \left( 1 + {1 \over 140} (1 -  \mu^2)( 17 - 5  \mu^2) C \right)
\ee
In this approximation flux function $\alpha$,
magnetic field components and twist angle are given by
\ba &&
\alpha = { \sqrt{2 C}   (1 -  \mu^2) \over  \tilde{r}  }
\nn &&
{B_r \over B_\ast} =  {  \mu \over  \tilde{r}^{3-8C/35}} \,
\left(1+ {3 (13 - 18 \mu^2 +5 \mu^4) C \over 140}\right)
\nn &&
{B_\theta  \over B_\ast} = { \sqrt{ 1 -  \mu^2}  \over 2 \tilde{r}^{3-8C/35}}
\, \left(1 - {(15 +  22 \mu^2 - 5 \mu^4) C \over 1400}\right)
\nn &&
{B_\phi \over B_\ast} = 
{ \sqrt{C}  (1 -  \mu^2)^{3/2} \over \sqrt{2}   \tilde{r}^2 }
\nn &&
\Delta \phi =  \sqrt{2 C}  \mu
\label{small}
\ea
These relations are useful for the order of magnitude estimates.
For example, for small twists the total poloidal current 
$I \sim \sin ^4 \theta / \tilde{r}$ is  concentrated 
near the  equator.

The simple twisted self-similar magnetosphere described above  is  not a  temporal sequence, but,  importantly,  it provides simple and clear order-of-magnitude estimates of the effects involved. Numerical solutions \citep{2012ApJ...754L..12P} are in a general agreement with analytical estimates. 

\section{Variations of the spin-down rate}

\subsection{The  CME model of magnetar bursts}

The processes that cause magnetar X-ray flares (and possibly the persistent
emission)  may be similar to those operating in the  Solar corona.
 The electric  currents within the start are 
slowly pushed into the magnetosphere, gated by {slow, plastic deformations} of
the neutron star crust.  This leads to gradual twisting of the magnetospheric
field lines, on time scales much longer than flares or burst, and
creates active magnetospheric regions similar to Sun's spots. Solutions described in Section \ref{force-free} have energy stored in the \ms\ that exceeds the energy of the dipolar configuration, Fig. \ref{EEdipole}, so that the dissipation of the supporting currents can power the magnetar high energy activity. 
 Initially, when the electric current (and possibly magnetic flux)
is pushed from the star into \ms,  the latter slowly adjusts to the changing boundary
conditions.  As more and more current is pushed into the magnetosphere
it eventually reaches a point of dynamical instability beyond which the stable
equilibrium can no longer be maintained. The loss of stability leads to a
rapid restructuring of magnetic configuration, on the \Alfven crossing time
scale, formation of narrow current sheets, and onset of magnetic dissipation.
As a result, a large amount of magnetic energy is converted into the bulk
motion kinetic energy and radiation. Moreover, the change of magnetic topology
allows formation of expanding magnetic loops that eventually break away from
the star \citep{lyut06,2008MNRAS.391..268G}.

\begin{figure}
\includegraphics[width=.99\linewidth]{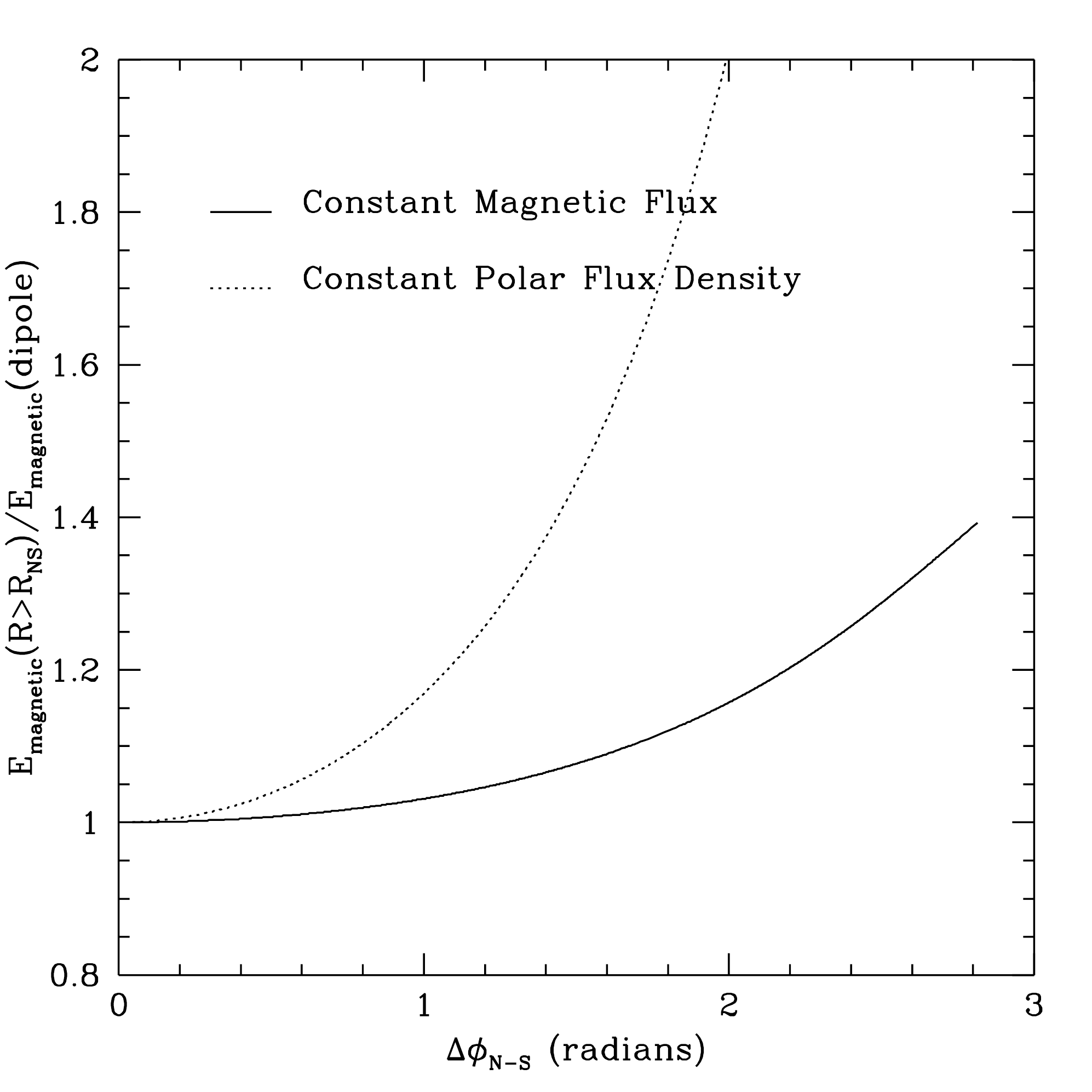}
\caption{ 
Energy contained in magnetosphere, compared with
 pure dipole field
}
\label{EEdipole}
\end{figure}

Adapting the 
 the generally accepted magnetic breakout model of
Solar Coronal Mass Ejections \citep{2008ApJ...683.1192L} to magnetar
environment, the underlying cause of all the 
manifestations of Solar activity - CMEs, eruptive flares and filament
ejections - is the disruption of a force balance between the upward pressure
of the strongly sheared field of a filament channel and the downward tension
of a potential (non-current carrying) overlying field.  Thus, an eruption is
driven solely by magnetic free energy stored in a closed, sheared magnetic
fields that open toward infinity during an eruption.    Reconnection  is thought to plays a
critical role in opening of \Bf\  lines: it removes the unsheared field above
the low-lying, sheared core flux, thereby allowing this core flux to burst
open.

\subsection{The magnetospheric  ``anti-glitch''}

Next we estimate parameter of the rapid spindown during  an $X$-ray flare.
Using the Goldrecih-Julian model, the spindown luminosity is related to the open magnetic flux $\Phi$: $L_{sd} \approx \Phi^2  \Omega^2 / c$.  In particular, for dipole \ms, $ \Phi \approx B_{NS}  R_{NS} ^2 ( \Omega R_{NS} /c)$. In the twisted \ms\ the open flux is larger \citep{tlk}, so that the 
spin-down rate is
\be
L_{sd} \sim B^2_\ast R_\ast^2 c \left( { \Omega  R_\ast \over c}\right)^{2(p+1)}, \, 0<p<1
\label{s}
\ee
Thus, 
\be
\dot{\Omega} \propto  \Omega ^{2p+1}, \, n = 2 p+1, 1<n<3
\ee
where $n$ is the spindown index.

The changes  of the magnetospheric structure can naturally induce large changes in the spindown rate. For example, 
for a given surface \Bf\ $B$  and pulsar spin frequency $\Omega$,  the maximal theoretical spindown rate (corresponding to the monopolar fields, $p=0, \, n=1$) is 
\ba &&
\dot{\Omega}_{max} = \eta_d^{-2} \dot{\Omega}_{d}  = 10^9 \dot{\Omega}_{d} 
\nn &&
\eta_d =  {   \Omega  r_\ast \over c }
\nn &&
\dot{\Omega}_{d} \sim{  B_{NS}^2 R_{NS}^2 c \over I_{NS} \Omega} \left( { \Omega  R_\ast \over c}\right)^4
\label{Wow}
\ea
where $\dot{\Omega}_{d}$ is a dipole spindown rate, $R_{NS}$ is \NS\ radius, $B_{NS}$ is surface \Bf, $ I_{NS}$ is the moment of inertia  and the numerical estimate is given for the particular magnetar.
Thus, in long period \NSs\ a relatively small change in the magnetospheric structure can induce huge variations in the spindown rate. This is due to the fact  that  in a dipolar-type configuration only  a tiny fraction of the field lines, of the order of $\eta_d  \sim 3 \times 10^{-5}$ contributes to the  spindown. A change in the fraction of the open flux then can induce large fluctuations of the spindown rate.

Let us assume 
that  the persistent spindown rate $\dot{\nu}_0 \approx - 10^{-14} $ Hz/s  \citep{2013Natur.497..591A} is determined mostly  by the dipolar-like spindown. Let the open magnetic flux during the outburst be a fraction $\eta$ of the total flux, so that the spindown rate is a faction $\eta^2 $ of the maximal value (\ref{Wow}), $\dot{\Omega} = \eta^2 \dot{\Omega}_{max}$. Then to produce an  ``anti-glitch'' of the order of $5 \times 10^{-8} $ Hz within one period it is required that the fraction of the open flux is 
\be
\eta \approx  \left( { \Omega  R_\ast \over c}\right) \sqrt{ {\Delta \nu \over \dot{\nu}_0 P}  } =  2.5 \times 10^{-2}
\label{eta}
\ee
Thus, all  that is required to produce an ``anti-glitch'' is to open only 
a few percent of the total magnetic flux through hemisphere. Or, in other words, the enhanced torque should be of the order of $10^{-4}-10^{-5}$ of the maximal possible torque acting during one rotational period. For longer active spindown periods, the requirement on $\eta$ is even smaller. These are very reasonable estimates. The amount of open magnetic flux during the flare is much larger than for dipole field, 
$
\eta \gg \eta_d
$, yet it is much smaller than the total magnetic flux through a hemisphere, $\eta \ll 1$. Using numerical simulations of the twisted \ms\ \citep{2012ApJ...754L..12P} came to similar conclusions, that opening a small fraction of the \ms\ can lead to rapid spindown.
The total energy involved in the opening of the \ms\  is a fraction $\sim \eta^2$ of the total \Bf\ energy. In case of 1E 2259+586 (\Bf\ of $6 \times 10^{13} $ G)  this amounts to $\sim 10^{41}$ ergs. This is consistent with   the energy released in the prompt GBM burst, $\sim 10^{37} $ erg, after accounting for the radiative efficiency of the resulting CME-like event. 
We conclude that the  ``anti-glitch'' behavior can be naturally explained as a transient high spindown rate induced by the magnetospheric changes.  

\subsection{Enhanced spindown and spectral hardening during $X$ activity}

Immediately following the $X$-ray burst, the magnetar  1E 2259+586 showed a period of enhanced activity during which the spindown rate was approximately 2-3 times higher than the average value. Below we show this this is due to a mild overall increase in the magnetospheric current. Let us use the analytical small current approximation (\ref{small}).
 In the self-similar model, for not very large twists, we have  $p=1-8/35 C, \, C = \Delta \phi^2/2$.
If magnetospheric structure changes (e.g., there is a jump in 
the current parameter $ \Delta C$ due to increased twist), this will induce a jump 
in the  spin-down parameter $ \Delta p = -8/35  \Delta C$ and
a corresponding jump in  spin-down rate.

%The estimate (\ref{Wow}) shows that in the long period \NSs\ like magnetars {\it it is very easy to change the spin down rate  by inducing relatively  mild   changes in  the magnetospheric structure}. (Note that the estimate (\ref{Wow}) compares the spindown rate of the twisted \ms\ with the dipolar spindown rate. The \ms\ in fact is distorted during the quiescent times as well, so,  even during quiescent time the spindown rate is enhanced with respect to the dipolar spin down. 

Let us assume that in a normal state a typical twist is $\Delta \phi_0$. As discussed above, it is required that spindown increased by a factor ${\cal N} \approx 2$ during the post burst phase.  This requires that during the activity time
\be
\Delta \phi ^2 = \Delta \phi_0  ^2 + {35 \over 16} {  \ln {\cal N}  \over \ln  R_{LC}/r_\ast}
\ee
Assuming $\Delta \phi ^2  \gg  \Delta \phi_0  ^2 $ and using the parameters of the magnetar and the required spin-down increase, we find
\be 
\Delta \phi  \approx 0.37
\label{twist}
\ee
This is a reasonable estimate: during the activity time (a  weeks after the $X$-ray flare  in  1E 2259+586)  the \ms\ on average should be twisted by about twenty degrees. 
Note, that  for large twists the spindown rate is a highly sensitive function of the twist: for a twenty degrees turn the spindown rate is 2 times the dipole, while for a full turn it is $10^9$ times the dipole, Eq. (\ref{Wow}). Using the estimate of the twist, and the small current solutions (\ref{small}), we
 find for the current parameter $C =   0.06$ and  $p =.98$, all within the assumed limit. 

The fact that the twist (\ref{twist}) is smaller than unity is important for the self-consistency of the model.
It is expected that highly twisted configurations are unstable towards kinks
or buckling instabilities.
Understanding
stability of the configurations  (\ref{sol})   
 remains an open question.

Note, that the spindown of a \NS\ on  long time scales (much longer than the period) are not uniquely related to the $X$-ray activity. The luminosity presumable comes from a large fraction of the \ms, mostly from closed field lines. At the same time, the spindown is determined by currents flowing along  a very small bundle of field lines. Thus we do not expect an exact one-to-one correspondence between the $X$-ray luminosity and the spindown rate, only a positive correlation.

On the other hand, suppose that 
  the crustal motion twists a patch of the surface by a given angle.
  % In the inhomogeneous \Bf\  the twist concentrates at a region  of  lowest field strength, \ie\ at the furthest point that a twisted flux tube reaches away from the star.
  Twisting the flux tube originating near the magnetic pole and extending nearly up to the light cylinder is more likely to lead to kink insatiability and opening of the \ms, since the instability of twisted closed   field lines is likely to be suppressed by the overlying dipolar fields. But twisting the fields near the magnetic pole strongly affects the  current responsible for the spindown of the star. Thus, it is natural to expect that a  burst or a flares lead to temporary  opening of the \ms\ and large increase in the  torque. 

\cite{2006MNRAS.368..690L} \citep[see also][]{2007ApJ...660..615F,2008ApJ...686.1245R,2013ApJ...762...13B} have constructed a model of resonant cyclotron scattering in the magnetar \mss\ that relates the spindown properties to the hardness of the soft non-thermal radiation.  According to the model, during the enhanced activity the \ms\ typically supports larger currents. Larger currents and larger particle density in the \ms\ increases the importance of resonant scattering, making the spectra harder.
The   optical depth
due to resonant scattering $\tau$  can be expressed in terms of the current parameter
$C$. For not very large twists
\be
 \tau \sim \sqrt{C} \propto \Delta \phi
\ee
(assuming  a constant velocity of scatters). The resulting spectrum is not a pure  power law, but is in agreement with the data. Qualitatively, the higher is the twist, the larger is the luminosity 
\citep[see Eq. (34) of ][]{tlk}, the harder is the spectrum.  

The total dissipated energy  is proportional to the current (and the twist),  $ L_X \propto \Delta \phi$, so that  $\tau \propto L_X$. Thus we expect that a substantial (order of unity) change in the $X$-ray luminosity is accompanied by a similar change in the spindown rate and a change in the spectral properties.
All these relations are  in agreement with the observations of  1E 2259+586.

\section{Discussion}

In this paper we showed that the transient (lasting for just one period)   opening of a relatively  small fraction of the \ms\ (a few percent) can explain the rapid spindown of the magnetar   1E 2259+586  observed by \cite{2013Natur.497..591A}. We point out that in magnetars it is very easy to have large variations of the spin-down rate - the theoretical upper limit is an enhancement by a factor up to $\sim 10^9$. 
 What is more, in the ensuing period of the enhanced activity the  behavior of the $X$-ray  luminosity, spindown rates  and spectral behavior  are  all consistent with the magnetospheric origin. We suggest, there may be  no real second glitch, only erratic behavior while the  magnetosphere relaxes to a long term steady state.

Magnetars are also known for their erratic timing behavior  \citep[\eg][]{2006csxs.book..547W}. The spindown rates can vary by a factor of few  over timescales of weeks \citep[eg][]{2002ApJ...576..381W}, while the giant flares are often accompanied by a sudden increase in the spin period \citep{1999ApJ...524L..55W}. Note that in case of magnetars the time ``resolution'' -  a typical interval between the measurements of the spin, are about a week.  This, combined with the fact that torque can fluctuate by as much as billion times, make it hard to define a glitch - as opposed to  smooth torque variation.

 In this paper we argue that timing fluctuations in magnetars are purely magnetospheric. \citep[In contrast, ][argued that  large  timing irregularities in magnetars may  have an internal origin, similar to glitches in  rotationally powered pulsars.]{2000ApJ...543..340T}. 
 We suggest that there are two types of glitches in magnetars:  (i)  due to topological changes of the \ms\ - opening up of a considerable magnetic flux (rapid spin down on the time scale of one rotation, associated with some radiative activity; this is specific to magnetars); (ii) smooth changes in the magnetospheric structure (radiatively silent, no large $X$-ray bursts, spindown changes can  be positive or negative; this is similar to the torque variations in the rotationally powered pulsars). In the second case there is a clear prediction: larger averaged $L_X$, larger spindown, harder spectra (this is generally consistent with the behavior of 1E 2259+586, as well as post-giant flare evolution, see discussion in   \cite{lyut06}).   The first case (opening up of the \ms) leads to transient   spindown events during bursts and flares and is specific to magnetars,  while the second  case (changes in the current flow in the \ms), produces fluctuations around some mean value and is analogues to torque fluctuations in rotationally-powered pulsar.

In comparison, in the rotationally powered pulsars 
there are two types of timing irregularities:   glitches, sudden changes of the spin frequency occurring on the time scale of single rotation period \citep[\eg][]{2011MNRAS.414.1679E} and long term spindown variations \citep{2006Sci...312..549K,2013Sci...339..436H}. Glitches are related to the internal \NS\ dynamics, how the superfluid vortexes interact with the crustal lattice \citep{2008LRR....11...10C}. The long term torque variations are assumed to be related to the changes in the magnetospheres current distribution, though a lack of general  understanding of  pulsar \mss\ impede  the development of  models \citep[see, though][]{2013MNRAS.429...20T}. We suggest that the second type, magnetospheric long term variations, is also operational in magnetars, though generally with larger spin variations. Magnetars also have new specific torque variation mechanism due to opening of the \ms.

We hypothesize that  in cases of so called spin-up glitches, the ``prompt'' short  spike, like the one detected by the GBM in   1E 2259+586,  was missed by the all sky monitors. Then the ensuing erratic behavior of the spin-down torque can lead to overcompensation of the prompt spin-down glitch.

As we discuss next, the overall timing behavior of  magnetars is in a general agreement with the magnetospheric model.
 Dib \& Kaspi (in preparation) point out that  radiative changes are almost always accompanied by some form of timing anomaly - in our model the magnetospheric opening is accompanied by a formation of a CME and associated prompt spike. Also, large long term magnetospheric changes should have correlated spindown and persistent $X$-ray flux increase. On the other hand,
only occasionally  timing anomalies in AXPs  are accompanied by any form of radiative anomaly. 
Most of the torque variations in our model come from the magnetospheric changes on the closed field lines -  such changes are  not necessarily expected   to be associated with large radiative effects. 
Importantly, the  changes in the magnetospheric structure of closed field lines should show in the changing pulse profiles. This agrees  with observations.

Superficially,  the timing glitches in normal pulsars and magnetars look similar:  roughly same amplitude,  similar recovery and inter-glitch times  (about a day or so).  We think these similarities are misleading. For example, the rate of accumulation of any mismatch between different components within the  \NS, $\propto \Omega^3$,  is different by an  order of a million between the  glitching pulsars (which are typically fast) and the slow magnetars. 

 There are other possible hints in favor of the magnetospheric origin of magnetar timing anomalies.  The magnetar  1E 1048.1-5937 has the highest rotational noise, has  one of the largest  flux variations and has the highest surface temperature (Dib \& Kaspi, in preparation). This is also consistent with the magnetospheric model: larger variations of the current are related to large flux and spindown variations. The high surface temperature is  related to higher overall magnetospheric currents. 
In addition,  the  timing noise is correlated with the strength of the \Bf: 1E 2259+586 and 4U 0142+61 have the least   noise and smallest \Bf, while 1RXS J1708-4009 and 1E 1841-045 are the opposite. 

Finally, there are magnetar activity events that seem to be  associated with spin-up glitches \citep[\eg\ the 2007 event in 1E 1048.1-5937,][]{2009ApJ...702..614D}. We note that at the time of the event there were very large variations in $\dot{\Omega}$, that, we hypothesize, when averaged over the interval between the observations produced an effective spin-up event.

I would like to thank  Robert Archibald,  Andrei Beloborodov, Konstantinos Gourgouliatos, Victoria Kaspi, Christopher Thompson and David Tsang for discussions and Department of Physics of  McGill University for hospitality. This research was partially supported by NASA  grant NNX12AF92G and
by  Simons Foundation. 

%\begin{figure}
%\includegraphics[width=.99\linewidth]{twist1e.pdf}
%\caption{ 
% The actual polar field inferred from spin parameters $P$ and $\dot P$,
%compared with the magnetic dipole formula.}
%\end{figure}

\appendix

\bibliographystyle{apj}
  \bibliography{/Users/maxim/Home/Research/BibTex}

\begin{thebibliography}{32}
\expandafter\ifx\csname natexlab\endcsname\relax\def\natexlab#1{#1}\fi

\bibitem[{{Archibald} {et~al.}(2013){Archibald}, {Kaspi}, {Ng},
  {Gourgouliatos}, {Tsang}, {Scholz}, {Beardmore}, {Gehrels}, \&
  {Kennea}}]{2013Natur.497..591A}
{Archibald}, R.~F., {Kaspi}, V.~M., {Ng}, C.-Y., {Gourgouliatos}, K.~N.,
  {Tsang}, D., {Scholz}, P., {Beardmore}, A.~P., {Gehrels}, N., \& {Kennea},
  J.~A. 2013, \nat, 497, 591

\bibitem[{{Beloborodov}(2013)}]{2013ApJ...762...13B}
{Beloborodov}, A.~M. 2013, \apj, 762, 13

\bibitem[{{Chamel} \& {Haensel}(2008)}]{2008LRR....11...10C}
{Chamel}, N. \& {Haensel}, P. 2008, Living Reviews in Relativity, 11, 10

\bibitem[{{Dib} {et~al.}(2009){Dib}, {Kaspi}, \&
  {Gavriil}}]{2009ApJ...702..614D}
{Dib}, R., {Kaspi}, V.~M., \& {Gavriil}, F.~P. 2009, \apj, 702, 614

\bibitem[{{Espinoza} {et~al.}(2011){Espinoza}, {Lyne}, {Stappers}, \&
  {Kramer}}]{2011MNRAS.414.1679E}
{Espinoza}, C.~M., {Lyne}, A.~G., {Stappers}, B.~W., \& {Kramer}, M. 2011,
  \mnras, 414, 1679

\bibitem[{{Fern{\'a}ndez} \& {Thompson}(2007)}]{2007ApJ...660..615F}
{Fern{\'a}ndez}, R. \& {Thompson}, C. 2007, \apj, 660, 615

\bibitem[{{Forbes} {et~al.}(2006){Forbes}, {Linker}, {Chen}, {Cid}, {K{\'o}ta},
  {Lee}, {Mann}, {Miki{\'c}}, {Potgieter}, {Schmidt}, {Siscoe}, {Vainio},
  {Antiochos}, \& {Riley}}]{ForbesCME}
{Forbes}, T.~G., {Linker}, J.~A., {Chen}, J., {Cid}, C., {K{\'o}ta}, J., {Lee},
  M.~A., {Mann}, G., {Miki{\'c}}, Z., {Potgieter}, M.~S., {Schmidt}, J.~M.,
  {Siscoe}, G.~L., {Vainio}, R., {Antiochos}, S.~K., \& {Riley}, P. 2006, Space
  Science Reviews, 123, 251

\bibitem[{{Gourgouliatos}(2008)}]{2008MNRAS.385..875G}
{Gourgouliatos}, K.~N. 2008, \mnras, 385, 875

\bibitem[{{Gourgouliatos} \& {Lynden-Bell}(2008)}]{2008MNRAS.391..268G}
{Gourgouliatos}, K.~N. \& {Lynden-Bell}, D. 2008, \mnras, 391, 268

\bibitem[{{Hermsen} {et~al.}(2013){Hermsen}, {Hessels}, {Kuiper}, {van
  Leeuwen}, {Mitra}, {de Plaa}, {Rankin}, {Stappers}, {Wright}, {Basu},
  {Alexov}, {Coenen}, {Grie{\ss}meier}, {Hassall}, {Karastergiou}, {Keane},
  {Kondratiev}, {Kramer}, {Kuniyoshi}, {Noutsos}, {Serylak}, {Pilia}, {Sobey},
  {Weltevrede}, {Zagkouris}, {Asgekar}, {Avruch}, {Batejat}, {Bell}, {Bell},
  {Bentum}, {Bernardi}, {Best}, {B{\^i}rzan}, {Bonafede}, {Breitling},
  {Broderick}, {Br{\"u}ggen}, {Butcher}, {Ciardi}, {Duscha}, {Eisl{\"o}ffel},
  {Falcke}, {Fender}, {Ferrari}, {Frieswijk}, {Garrett}, {de Gasperin}, {de
  Geus}, {Gunst}, {Heald}, {Hoeft}, {Horneffer}, {Iacobelli}, {Kuper}, {Maat},
  {Macario}, {Markoff}, {McKean}, {Mevius}, {Miller-Jones}, {Morganti}, {Munk},
  {Orr{\'u}}, {Paas}, {Pandey-Pommier}, {Pandey}, {Pizzo}, {Polatidis},
  {Rawlings}, {Reich}, {R{\"o}ttgering}, {Scaife}, {Schoenmakers}, {Shulevski},
  {Sluman}, {Steinmetz}, {Tagger}, {Tang}, {Tasse}, {ter Veen}, {Vermeulen},
  {van de Brink}, {van Weeren}, {Wijers}, {Wise}, {Wucknitz}, {Yatawatta}, \&
  {Zarka}}]{2013Sci...339..436H}
{Hermsen}, W., {Hessels}, J.~W.~T., {Kuiper}, L., {van Leeuwen}, J., {Mitra},
  D., {de Plaa}, J., {Rankin}, J.~M., {Stappers}, B.~W., {Wright}, G.~A.~E.,
  {Basu}, R., {Alexov}, A., {Coenen}, T., {Grie{\ss}meier}, J.-M., {Hassall},
  T.~E., {Karastergiou}, A., {Keane}, E., {Kondratiev}, V.~I., {Kramer}, M.,
  {Kuniyoshi}, M., {Noutsos}, A., {Serylak}, M., {Pilia}, M., {Sobey}, C.,
  {Weltevrede}, P., {Zagkouris}, K., {Asgekar}, A., {Avruch}, I.~M., {Batejat},
  F., {Bell}, M.~E., {Bell}, M.~R., {Bentum}, M.~J., {Bernardi}, G., {Best},
  P., {B{\^i}rzan}, L., {Bonafede}, A., {Breitling}, F., {Broderick}, J.,
  {Br{\"u}ggen}, M., {Butcher}, H.~R., {Ciardi}, B., {Duscha}, S.,
  {Eisl{\"o}ffel}, J., {Falcke}, H., {Fender}, R., {Ferrari}, C., {Frieswijk},
  W., {Garrett}, M.~A., {de Gasperin}, F., {de Geus}, E., {Gunst}, A.~W.,
  {Heald}, G., {Hoeft}, M., {Horneffer}, A., {Iacobelli}, M., {Kuper}, G.,
  {Maat}, P., {Macario}, G., {Markoff}, S., {McKean}, J.~P., {Mevius}, M.,
  {Miller-Jones}, J.~C.~A., {Morganti}, R., {Munk}, H., {Orr{\'u}}, E., {Paas},
  H., {Pandey-Pommier}, M., {Pandey}, V.~N., {Pizzo}, R., {Polatidis}, A.~G.,
  {Rawlings}, S., {Reich}, W., {R{\"o}ttgering}, H., {Scaife}, A.~M.~M.,
  {Schoenmakers}, A., {Shulevski}, A., {Sluman}, J., {Steinmetz}, M., {Tagger},
  M., {Tang}, Y., {Tasse}, C., {ter Veen}, S., {Vermeulen}, R., {van de Brink},
  R.~H., {van Weeren}, R.~J., {Wijers}, R.~A.~M.~J., {Wise}, M.~W., {Wucknitz},
  O., {Yatawatta}, S., \& {Zarka}, P. 2013, Science, 339, 436

\bibitem[{{Horowitz} {et~al.}(2011){Horowitz}, {Hughto}, {Schneider}, \&
  {Berry}}]{2011arXiv1109.5095H}
{Horowitz}, C.~J., {Hughto}, J., {Schneider}, A., \& {Berry}, D.~K. 2011, ArXiv
  e-prints

\bibitem[{{Jones}(2003)}]{2003ApJ...595..342J}
{Jones}, P.~B. 2003, \apj, 595, 342

\bibitem[{{Komissarov} {et~al.}(2007){Komissarov}, {Barkov}, \&
  {Lyutikov}}]{2007MNRAS.374..415K}
{Komissarov}, S.~S., {Barkov}, M., \& {Lyutikov}, M. 2007, \mnras, 374, 415

\bibitem[{{Kramer} {et~al.}(2006){Kramer}, {Lyne}, {O'Brien}, {Jordan}, \&
  {Lorimer}}]{2006Sci...312..549K}
{Kramer}, M., {Lyne}, A.~G., {O'Brien}, J.~T., {Jordan}, C.~A., \& {Lorimer},
  D.~R. 2006, Science, 312, 549

\bibitem[{{Lynch} {et~al.}(2008){Lynch}, {Antiochos}, {DeVore}, {Luhmann}, \&
  {Zurbuchen}}]{2008ApJ...683.1192L}
{Lynch}, B.~J., {Antiochos}, S.~K., {DeVore}, C.~R., {Luhmann}, J.~G., \&
  {Zurbuchen}, T.~H. 2008, \apj, 683, 1192

\bibitem[{{Lynden-Bell} \& {Boily}(1994)}]{1994MNRAS.267..146L}
{Lynden-Bell}, D. \& {Boily}, C. 1994, \mnras, 267, 146

\bibitem[{{Lyutikov}(2003)}]{LyutikovTear}
{Lyutikov}, M. 2003, \mnras, 346, 540

\bibitem[{{Lyutikov}(2006)}]{lyut06}
---. 2006, \mnras, 367, 1594

\bibitem[{{Lyutikov} \& {Gavriil}(2006)}]{2006MNRAS.368..690L}
{Lyutikov}, M. \& {Gavriil}, F.~P. 2006, \mnras, 368, 690

\bibitem[{{Palmer} {et~al.}(2005){Palmer}, {Barthelmy}, {Gehrels}, {Kippen},
  {Cayton}, {Kouveliotou}, {Eichler}, {Wijers}, {Woods}, {Granot}, {Lyubarsky},
  {Ramirez-Ruiz}, {Barbier}, {Chester}, {Cummings}, {Fenimore}, {Finger},
  {Gaensler}, {Hullinger}, {Krimm}, {Markwardt}, {Nousek}, {Parsons}, {Patel},
  {Sakamoto}, {Sato}, {Suzuki}, \& {Tueller}}]{palmer}
{Palmer}, D.~M., {Barthelmy}, S., {Gehrels}, N., {Kippen}, R.~M., {Cayton}, T.,
  {Kouveliotou}, C., {Eichler}, D., {Wijers}, R.~A.~M.~J., {Woods}, P.~M.,
  {Granot}, J., {Lyubarsky}, Y.~E., {Ramirez-Ruiz}, E., {Barbier}, L.,
  {Chester}, M., {Cummings}, J., {Fenimore}, E.~E., {Finger}, M.~H.,
  {Gaensler}, B.~M., {Hullinger}, D., {Krimm}, H., {Markwardt}, C.~B.,
  {Nousek}, J.~A., {Parsons}, A., {Patel}, S., {Sakamoto}, T., {Sato}, G.,
  {Suzuki}, M., \& {Tueller}, J. 2005, \nat, 434, 1107

\bibitem[{{Parfrey} {et~al.}(2012){Parfrey}, {Beloborodov}, \&
  {Hui}}]{2012ApJ...754L..12P}
{Parfrey}, K., {Beloborodov}, A.~M., \& {Hui}, L. 2012, \apjl, 754, L12

\bibitem[{{Pavan} {et~al.}(2009){Pavan}, {Turolla}, {Zane}, \&
  {Nobili}}]{2009MNRAS.395..753P}
{Pavan}, L., {Turolla}, R., {Zane}, S., \& {Nobili}, L. 2009, \mnras, 395, 753

\bibitem[{{Rea} {et~al.}(2008){Rea}, {Zane}, {Turolla}, {Lyutikov}, \&
  {G{\"o}tz}}]{2008ApJ...686.1245R}
{Rea}, N., {Zane}, S., {Turolla}, R., {Lyutikov}, M., \& {G{\"o}tz}, D. 2008,
  \apj, 686, 1245

\bibitem[{{Shafranov}(1966)}]{1966RvPP....2..103S}
{Shafranov}, V.~D. 1966, Reviews of Plasma Physics, 2, 103

\bibitem[{{Thompson} \& {Duncan}(1995)}]{TD95}
{Thompson}, C. \& {Duncan}, R.~C. 1995, \mnras, 275, 255

\bibitem[{{Thompson} {et~al.}(2000){Thompson}, {Duncan}, {Woods},
  {Kouveliotou}, {Finger}, \& {van Paradijs}}]{2000ApJ...543..340T}
{Thompson}, C., {Duncan}, R.~C., {Woods}, P.~M., {Kouveliotou}, C., {Finger},
  M.~H., \& {van Paradijs}, J. 2000, \apj, 543, 340

\bibitem[{{Thompson} {et~al.}(2002){Thompson}, {Lyutikov}, \& {Kulkarni}}]{tlk}
{Thompson}, C., {Lyutikov}, M., \& {Kulkarni}, S.~R. 2002, \apj, 574, 332

\bibitem[{{Timokhin} \& {Arons}(2013)}]{2013MNRAS.429...20T}
{Timokhin}, A.~N. \& {Arons}, J. 2013, \mnras, 429, 20

\bibitem[{{Wolfson}(1995)}]{1995ApJ...443..810W}
{Wolfson}, R. 1995, \apj, 443, 810

\bibitem[{{Woods} {et~al.}(2002){Woods}, {Kouveliotou}, {G{\"o}{\v g}{\"u}{\c
  s}}, {Finger}, {Swank}, {Markwardt}, {Hurley}, \& {van der
  Klis}}]{2002ApJ...576..381W}
{Woods}, P.~M., {Kouveliotou}, C., {G{\"o}{\v g}{\"u}{\c s}}, E., {Finger},
  M.~H., {Swank}, J., {Markwardt}, C.~B., {Hurley}, K., \& {van der Klis}, M.
  2002, \apj, 576, 381

\bibitem[{{Woods} {et~al.}(1999){Woods}, {Kouveliotou}, {van Paradijs},
  {Finger}, {Thompson}, {Duncan}, {Hurley}, {Strohmayer}, {Swank}, \&
  {Murakami}}]{1999ApJ...524L..55W}
{Woods}, P.~M., {Kouveliotou}, C., {van Paradijs}, J., {Finger}, M.~H.,
  {Thompson}, C., {Duncan}, R.~C., {Hurley}, K., {Strohmayer}, T., {Swank}, J.,
  \& {Murakami}, T. 1999, \apjl, 524, L55

\bibitem[{{Woods} \& {Thompson}(2006)}]{2006csxs.book..547W}
{Woods}, P.~M. \& {Thompson}, C. {Soft gamma repeaters and anomalous X-ray
  pulsars: magnetar candidates}, ed. W.~H.~G. {Lewin} \& M.~{van der Klis},
  547--586

\end{thebibliography}

\appendix

\section{$p\rightarrow 0$ limit}
\label{p=0}

Interesting analytical approximations can be obtained in the
limit $p \rightarrow 0$. In this case function $F$ is almost
a straight line with a slope $-2$ everywhere except near 
the point $\mu=0$ with  $F(0) \sim 2$. 
Near this point we can neglect $\mu^2$.
Integrating  Eq. (\ref{gseq}) once we find
\be 
p (p+1) F^2 + { C p \over p+1} F^{2 (p+1)/p} + {F'}^2=4
\ee
where constant of integration has been chose so that at $\mu=1$,
where $F=0$, ${F'}^2=4$.
 Since at  $\mu \rightarrow 0 $ the function $F  \sim 2 $ is raised
to a large power we can neglect the term $p(p+1)F $ as compared with
$ C F^{(p+2)/p}$:
\be 
{  C p \over p+1} F^{2 (p+1)/p} + {F'}^2=4
\ee
Making a substitution
\be
F \rightarrow \left({ p C \over 4 (p+1)} \right)^{ - {p \over 2(p+1)}}
\left(1 - { p \over p+1} \ln q \right)\equiv
{ 2 (p+1) \over C_2 p} \left(1 - { p \over p+1} \ln q \right),
\ee
the 
equation for the function $q$ reads
\be 
{q' \over \sqrt{q^2-1}} =C_2
\label{q}
\ee
where
\be
C_2 = \left({ 2 (p+1) \over p} \right)^{1/(p+1)}
 \left( { C (p+1) \over p} \right)^{p/(2 (p+1))}
\ee
Boundary conditions for $q$ are
\ba &
1 - { p \over p+1} \ln q=0  & \mbox{ at $\mu=1$}
\nn &
q'=0 & \mbox{ at $\mu=0$}
\ea
(two boundary conditions for the first order ODE (\ref{q}) are
needed to find the solution and dependence $p(C)$).
Eq. (\ref{q}) has a solution
\be
q= \cosh(C_2 \mu +C_1)
\ee
Boundary conditions give 
\ba & 
C_1=0 &
\nn &
C_2 = \cosh^{-1} \exp^{(p+1)/p} \sim (p+1)/p + \ln 2 &
\ea
which  resolves implicitly  $p(C)$
\be
C= { 2^{-2/p} (p+1)  \over p}
\ee
and determines the  flux function
\ba &&
F= { 2 \over   \cosh^{-1} \exp^{(p+1)/p}} 
\, \left( {p+1\over p} - \ln \cosh(\mu  \cosh^{-1} \exp^{(p+1)/p}) \right)
\nn &&
\sim { 2 (p+1) \over
1+p+ p \ln 2}  \left( 1 - { p\over p+1} \ln \cosh \mu (1+1/p+\ln2)  \right).
\ea
\citep[compare with][]{1994MNRAS.267..146L}.
Value of $F$ at $\mu=0$ is then
\be
F(0) = { 2 (p+1) \over 
1+p+ p \ln 2} \sim 2 (1- p \ln 2)
\ee
%These 
%asymptotic analytical relations are well approximated by numerical
%results.

\end{document}